\documentclass[a4paper,twocolumn,final,oneside,journal]{IEEEtran}

\IEEEoverridecommandlockouts

\usepackage{tikz}
\usetikzlibrary{
	arrows.meta,
	positioning,
	shapes.geometric,
	shapes.symbols,
	calc,
	fit,
	backgrounds,
	decorations.pathreplacing
}
\usepackage[american]{circuitikz}
\usepackage{epstopdf}
\usepackage{graphicx}
\usepackage{float}
\usepackage{pict2e}
\graphicspath{{./}{imgs/}{Figs/}{figs/}}

\usepackage{steinmetz}
\usepackage{amsmath}
\usepackage{amssymb}
\usepackage{stackrel}
\usepackage{icomma}
\usepackage{mathtools}

\usepackage[T1]{fontenc}
\usepackage[utf8]{inputenc}
\usepackage[main=english,brazil]{babel}
\usepackage{color}
\usepackage{soul,xcolor}
\usepackage{wasysym}
\usepackage{comment}
\usepackage{lineno}
\usepackage{stfloats}

\usepackage{multirow, bigdelim}
\usepackage{threeparttable}
\usepackage{booktabs}
\usepackage{longtable}
\usepackage{array}
\usepackage{tabularx}
\usepackage{makecell}

\usepackage{cite}
\usepackage{hyperref}
\hypersetup{
	colorlinks=true,
	linkcolor=black,
	filecolor=magenta,
	urlcolor=cyan,
	citecolor=black
}

\makeatletter
\ctikzset{lx/.code args={#1 and #2}{
  \pgfkeys{/tikz/circuitikz/bipole/label/name=\parbox{1cm}{\centering #1  \\ #2}}
    \ctikzsetvalof{bipole/label/unit}{}
    \ifpgf@circ@siunitx
        \pgf@circ@handleSI{#2}
        \ifpgf@circ@siunitx@res
            \edef\pgf@temp{\pgf@circ@handleSI@val}
            \pgfkeyslet{/tikz/circuitikz/bipole/label/name}{\pgf@temp}
            \edef\pgf@temp{\pgf@circ@handleSI@unit}
            \pgfkeyslet{/tikz/circuitikz/bipole/label/unit}{\pgf@temp}
        \else
        \fi
    \else
    \fi
}}

\ctikzset{lx^/.style args={#1 and #2}{
    lx=#2 and #1,
    \circuitikzbasekey/bipole/label/position=90 }
}

\ctikzset{lx_/.style args={#1 and #2}{
    lx=#1 and #2,
    \circuitikzbasekey/bipole/label/position=-90 }
}
\makeatother

\makeatletter
\newcommand\doslashcirc[2]{
	\sbox\z@{$#1\m@th\circ$}
	\setlength\unitlength{\wd\z@}
	\begin{picture}(1,1)
	\roundcap
	\put(0,0){\box\z@}
	\put(0,0){\line(1,1){1}}
	\end{picture}
}
\makeatother

\newcolumntype{L}[1]{>{\raggedright\arraybackslash}p{#1}}
\newcolumntype{C}[1]{>{\centering\arraybackslash}p{#1}}
\newcolumntype{Y}[1]{>{\raggedright\arraybackslash}p{#1}}

\tikzset{
	etapa/.style={
		signal, signal to=east, signal from=west,
		signal pointer angle=140,
		draw=black!55, line width=0.4pt,
		minimum height=1.0cm, text width=#1, align=center,
		inner xsep=3pt, font=\scriptsize
	},
	etapa/.default=1.8cm,
	primeira/.style={signal from=nowhere},
	cEntrada/.style={fill=orange!35},
	cKrylov/.style={fill=green!25},
	cSub/.style={fill=green!45},
	cProj/.style={fill=teal!30},
	cSaida/.style={fill=blue!25},
	rotulo/.style={font=\small\itshape, anchor=west}
}

\definecolor{evdc}{RGB}{70,130,180}   
\definecolor{lzc}{RGB}{255,140,0}     

\newcommand{\herm}{^{\mathsf{H}}}
\newcommand{\trans}{\mathsf{T}}

\newcommand{\bigO}{\mathcal{O}}
\newcommand{\Kry}{\mathcal{K}}
\newcommand{\ju}{\mathrm{j}}
\newcommand{\mA}{\mathbf{A}}
\newcommand{\mF}{\mathbf{F}}
\newcommand{\mI}{\mathbf{I}}
\newcommand{\mK}{\mathbf{K}}
\newcommand{\mQ}{\mathbf{Q}}
\newcommand{\mR}{\mathbf{R}}
\newcommand{\mS}{\mathbf{S}}
\newcommand{\mT}{\mathbf{T}}
\newcommand{\mU}{\mathbf{U}}
\newcommand{\mV}{\mathbf{V}}
\newcommand{\mX}{\mathbf{X}}
\newcommand{\mN}{\mathbf{N}}
\newcommand{\mLam}{\boldsymbol{\Lambda}}

\newcommand{\mRz}{\mathbf{R}_{\mathrm{z}}}
\newcommand{\mUs}{\mathbf{U}_{\mathrm{s}}}
\newcommand{\mUn}{\mathbf{U}_{\mathrm{n}}}
\newcommand{\mPn}{\mathbf{P}_{\mathrm{n}}}

\newcommand{\va}{\mathbf{a}}
\newcommand{\vv}{\mathbf{v}}

\newcommand{\vr}{\mathbf{r}}

\newcommand{\mRhat}{\hat{\mathbf{R}}}
\newcommand{\Loss}{\mathcal{L}}
\newcommand{\vu}{\mathbf{u}}
\newcommand{\vtheta}{\boldsymbol{\theta}}
\newcommand{\Odeg}{^{\circ}}
\DeclareMathOperator{\Span}{span}
\DeclareMathOperator{\diag}{diag}

\newcommand{\mTm}{\mathbf{T}_m}

\begin{document}


\ctikzset{
	american resistors,
	american inductors,
	full diodes,
	bipoles/cuteswitch/thickness=0.3,
	diodes/scale=0.5,
	resistors/scale=0.7,
	capacitors/scale=0.7,
	capacitors/thickness=4,
}

\title{Modified CNN-Learned Surrogate Covariance With Krylov/Lanczos Subspace for DoA Estimation}

\makeatletter
\newcommand{\linebreakand}{
	\end{@IEEEauthorhalign}
	\hfill\mbox{}\par
	\mbox{}\hfill\begin{@IEEEauthorhalign}
}
\makeatother

\author{
	\IEEEauthorblockN{
		R. Brandt$^1$,
		F. Rangel de Sousa$^2$,~\IEEEmembership{Senior Member,~IEEE},
		V. J. Brusamarello$^1$,~\IEEEmembership{Senior Member,~IEEE}
	}

	\IEEEauthorblockA{
		\textit{$^1$Universidade Federal do Rio Grande do Sul} UFRGS, Porto Alegre, Brasil - PPGEE \\
		\textit{$^2$Universidade Federal de Santa Catarina}
		UFSC, Florianópolis, Brasil - PPGEEL\\
		rodrigo.brandt@ufrgs.br, fernando.rangel.sousa@ufsc.br, valner.brusamarello@ufrgs.br
	}
}

\maketitle

\begin{abstract}
	Recently, learning-based approaches have been employed to improve the robustness of subspace methods in DoA estimation. A Convolutional Neural Network (CNN) was used as a data-driven framework employing an autoencoder to generate a well-conditioned surrogate covariance optimized end-to-end through a differentiable classical estimator, such as the Root-MUSIC. The complete system is designed to estimate the directions of arrival based on a spectral decomposition using the EVD of this covariance matrix. In this work, a new approach is proposed by formulating the spectral stage of the Root-MUSIC algorithm in terms of the orthogonal complement of the signal subspace, thereby eliminating the need for the noise subspace and noise eigenvalues. Furthermore, the full EVD is replaced by a Lanczos tridiagonalization with reorthogonalization, reducing the overall computational cost.
\end{abstract}

\begin{IEEEkeywords}
	CNN; hybrid method; DoA; Krylov, Lanczos, spectral decomposition.
\end{IEEEkeywords}

\IEEEpeerreviewmaketitle

\section{Introduction}
\label{sec:intro}

Direction-of-arrival (DoA) estimation is a fundamental problem in array signal processing, with applications in radar, sonar, and wireless communications~\cite{526899}. A common approach considers an array of $N$ sensors receiving signals from $M$ impinging sources and the signals observed over $T$ \textit{snapshots}, with $M<N$. Subspace-based methods~\cite{11180969}, such as MUSIC~\cite{1143830}, Root-MUSIC~\cite{1172124}, and ESPRIT~\cite{32276}, can achieve good angular resolution in DoA estimation by separating the observation space into orthogonal signal and noise subspaces, by eigenvalue decomposition (EVD) of the sample covariance matrix. Their accuracy, however, relies on restrictive assumptions, such as the presence of noncoherent sources, a sufficient number of \textit{snapshots}, and a properly calibrated array.

Such assumptions have motivated the development of learning-based approaches to improve the robustness of subspace methods under adverse operating conditions \cite{10750453,10748418}. For instance, SubspaceNet~\cite{10750453} introduces a data-driven framework with a Convolutional neural network (CNN). Instead of relying directly on the sample covariance matrix, SubspaceNet employs a convolutional autoencoder to generate a well-conditioned covariance surrogate with the Root-MUSIC estimator. Since no ground-truth covariance matrix is available as a training target, the network is trained according to the quality of the resulting DoA estimates, with the angular estimation error evaluated at the output of the estimator~\cite{merkofer2022damusic}.


However, once the surrogate covariance matrix is built, an eigenvalue decomposition (EVD) is still required to extract the signal and noise subspaces.
The EVD of an $N\times N$ matrix has a $\bigO(N^3)$ complexity and is performed once per batch sample, per epoch. In a typical training setup ($4096$ samples $\times$ $50$ epochs, corresponding to approximately $2\times10^5$ calls), the difference between $5$~ms and $15$~s per call turns a training process that would take minutes into one that takes weeks. Also, EVD process must be backpropagated through and closely spaced eigenvalues can lead to diverging gradients. \cite{wang2019backpropagationfriendlyeigendecomposition}.

Krylov methods have been employed for decades to reduce the dimensionality of large-scale problems, both in reduced-order modeling~\cite{Bai2002} and in fast decomposition techniques for real-time signal processing~\cite{Wu2015}. In the context of DoA estimation, Krylov subspace methods have been applied in angle and frequency estimation~\cite{4775696}, combined with sparse regularization and weighted $l_1$-norm formulations~\cite{10433433,9415040}, to reduce the computational cost of ESPRIT~\cite{6075507}  and adaptive subspace tracking~\cite{365290}. In this work, we propose the integration of a Krylov subspace method into a differentiable subspace estimation. The proposed approach substantially reduces the computational cost of subspace extraction while preserving, and in some cases improving, the estimation performance of conventional approaches. Furthermore, the proposed formulation exhibits good numerical stability, making it suitable for efficient subspace-based DoA estimation.

The main contributions of this work include the formulation of Root-MUSIC algorithm in terms of the orthogonal complement of the signal subspace, eliminating the need for the noise subspace and its associated eigenvalues. Furthermore, the full EVD is replaced by a Krylov subspace method (Lanczos algorithm), reducing the computational complexity from $\bigO(N^3)$ to $\bigO(mN^2)$ for fixed $m$. In this scenario, the backpropagation run through the eigendecomposition of a small real symmetric tridiagonal matrix. Finally, the proposed approach was validated over more than one hundred configurations involving $N$ sensors, signal-to-noise ratio (SNR), T \emph{snapshots}, Krylov subspace dimension, and source coherence. The results show Root Mean Square Periodic Error (RMSPE) values matching to four decimal places, speedups of up to $289\times$ for $N=2048$, and a hybrid diagnostic that isolates and eliminates the effect of inference.

\section{Signal Model and the CNN-based subspace method}
\label{sec:modelo}

\subsection{Signal Observation Model}
Consider the problem of estimating the DoA of $M$ signals impinging on a uniform linear array (ULA) of $N$ sensors (Fig.~\ref{fig:model}a). The observation model is given by
\begin{equation}
	\label{eq:modelo}
	\mX = \mA(\vtheta)\,\mS + \mN,
\end{equation}
where $\mX\in\mathbb{C}^{N\times T}$ is the observation matrix, with each column corresponding to one \emph{snapshot}, $\mA(\vtheta)\in\mathbb{C}^{N\times M}$ is the steering matrix whose columns $\va(\theta_i)$ depend on the unknown DoAs $\vtheta=[\theta_1,\dots,\theta_M]^\trans$, $\mS\in\mathbb{C}^{M\times T}$ contains the source signals, and $\mN\in\mathbb{C}^{N\times T}$ denotes additive noise~\cite{526899}. Classical subspace-based methods estimate $\vtheta$ from the eigendecomposition (EVD) of the covariance matrix
$\mR=\mathbb{E}[\mX\mX\herm]$, which in practice is approximated by the sample covariance matrix
$\widehat{\mR}=\mX\mX\herm/T\in\mathbb{C}^{N\times N}$.
The EVD of $\widehat{\mR}$ is given by
$\widehat{\mR}\mU=\mU\mLam$, where $\mU\in\mathbb{C}^{N\times N}$ contains the eigenvectors and $\mLam\in\mathbb{R}^{N\times N}$ is a diagonal matrix containing the corresponding eigenvalues. Assuming the eigenvalues are ordered in descending order, the first $M$ eigenvectors span the \emph{signal subspace}, whereas the remaining $N-M$ eigenvectors span the \emph{noise subspace}. The orthogonality between the steering vectors $\va(\theta)$ and the noise subspace is then exploited to identify the true DoAs.

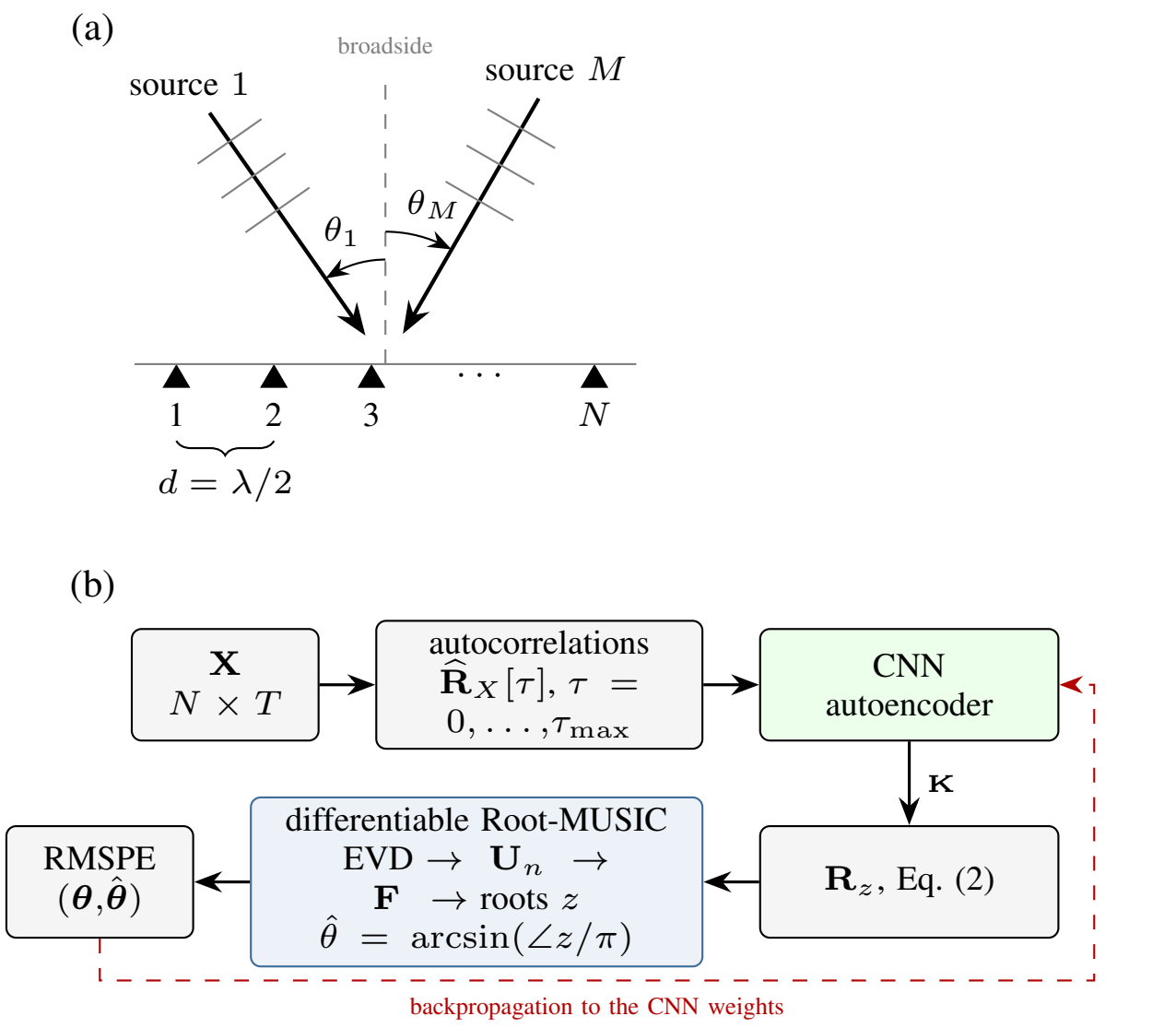
\begin{figure}[!t]
	\centering
	\begin{tikzpicture}[font=\scriptsize, >=Stealth]
		\foreach \i/\x in {1/0,2/0.7,3/1.4} {
				\fill (\x,0) -- ++(-0.1,-0.17) -- ++(0.2,0) -- cycle;
				\node[below=1.5mm] at (\x,0) {\i};
			}
		\node at (2.2,-0.08) {$\cdots$};
		\fill (3.0,0) -- ++(-0.1,-0.17) -- ++(0.2,0) -- cycle; \node[below=1.5mm] at (3.0,0) {$N$};
		\draw[gray] (-0.3,0) -- (3.3,0);
		\draw[decorate, decoration={brace, amplitude=3pt, mirror}] (0,-0.55) -- (0.7,-0.55)
		node[midway, below=2pt] {$d=\lambda/2$};
		\coordinate (c) at (1.5,0);
		\draw[dashed, gray] (c) -- ++(0,2.1) node[above, font=\tiny] {broadside};
		\foreach \th/\lab/\nm in {-35/1/1, 30/M/M} {
				\draw[->, thick] ($(c)+(90-\th:2.2)$) -- ($(c)+(90-\th:0.25)$);
				\foreach \r in {1.35,1.65,1.95}
				\draw[gray] ($(c)+(90-\th:\r)+(-\th:0.28)$) -- ($(c)+(90-\th:\r)+(180-\th:0.28)$);
				\node at ($(c)+(90-\th:2.45)$) {source $\nm$};
			}
		\draw[->] ($(c)+(90:0.75)$) arc (90:125:0.75); \node at ($(c)+(108:1.0)$) {$\theta_1$};
		\draw[->] ($(c)+(90:0.95)$) arc (90:60:0.95);  \node at ($(c)+(73:1.2)$) {$\theta_M$};
		\node[font=\footnotesize] at (-0.6,2.4) {(a)};

		\begin{scope}[yshift=-1.75cm,
				blk/.style={draw, rounded corners=2pt, align=center, inner sep=2pt, minimum height=8mm, fill=gray!8},
				arr/.style={->, semithick}]
			\node[font=\footnotesize] at (-0.6,0.15) {(b)};
			\node[blk, text width=12mm] (x)   at (0.35,-0.55) {$\mathbf{X}$\\ $N\times T$};
			\node[blk, text width=22mm, right=4mm of x] (ac) {autocorrelations\\ $\widehat{\mathbf{R}}_X[\tau]$, $\tau=0,\ldots,\tau_{\max}$};
			\node[blk, text width=20mm, right=4mm of ac, fill=green!8] (cnn) {CNN\\ autoencoder};
			\node[blk, text width=20mm, below=6mm of cnn] (rz) {$\mathbf{R}_z$, Eq.~(2)};
			\node[blk, text width=31mm, left=4mm of rz, fill=evdc!10, draw=evdc!85!black] (rm)
			{differentiable Root-MUSIC\\ EVD $\rightarrow\mathbf{U}_n\rightarrow\mathbf{F}\rightarrow$ roots $z$\\ $\hat\theta=\arcsin(\angle z/\pi)$};
			\node[blk, text width=12mm, left=4mm of rm] (loss) {RMSPE\\ $(\boldsymbol\theta,\hat{\boldsymbol\theta})$};
			\draw[arr] (x) -- (ac); \draw[arr] (ac) -- (cnn);
			\draw[arr] (cnn) -- node[right, font=\tiny] {$\mathbf{K}$} (rz);
			\draw[arr] (rz) -- (rm); \draw[arr] (rm) -- (loss);
			\draw[->, dashed, red!70!black] (loss.south) -- ++(0,-0.3) -| node[pos=0.25, below, font=\tiny] {backpropagation to the CNN weights}
			($(cnn.east)+(0.25,0)$) -- (cnn.east);
		\end{scope}

	\end{tikzpicture}

	\caption{(a) Uniform linear array of $N$ sensors with spacing $d=\lambda/2$, receiving $M$ far-field narrowband plane waves from directions $\theta_1,\ldots,\theta_M$ measured from broadside. (b) SubspaceNet [6]: the autocorrelation tensor of the snapshots is mapped by a convolutional autoencoder to $\mathbf{K}$, which defines the surrogate covariance $\mathbf{R}_z$ in (2); a differentiable Root-MUSIC returns $\hat{\boldsymbol\theta}$, and the RMSPE between $\hat{\boldsymbol\theta}$ and $\boldsymbol\theta$ is backpropagated to the network weights (dashed).}
	\label{fig:model}
\end{figure}

\subsection{CNN-based subspace method}

CNN-based subspace methods, such as SubspaceNet~\cite{10750453} replaces the raw covariance matrix with a learned \emph{surrogate} covariance matrix and incorporate a Root-MUSIC estimation procedure with gradient backpropagation to train the neural network (Fig.~\ref{fig:model}b).

The input to SubspaceNet is a tensor of sample autocorrelation matrices computed at multiple lags $\widehat{R}_X[\tau]$. An encoder-decoder processes this input and produces a complex-valued matrix $\mK\in\mathbb{C}^{N\times N}$. The network is trained end-to-end by incorporating the differentiable Root-Music algorithm into the training loop, allowing the DoA estimation error to be backpropagated to the network parameters. The learned matrix $\mathbf{K}$ defines the surrogate covariance matrix
\begin{equation}
	\mathbf R_z=\mathbf K\mathbf K^H+\epsilon\mathbf I_N,
	\label{eq:surrogate}
\end{equation}
where $\epsilon>0$ is a regularization constant and $\mathbf R_z$ is Hermitian and positive definite. Finally the root-MUSIC algorithm operates  on $R_z$ performing eigendecomposition to obtain the noise subspace $U_n$, subspace projector $\mathbf F=\mathbf U_n\mathbf U_n^H$ and the polynomial coefficients are derived. The selected signal roots are subsequently used to estimate DoAs, acoording to
\begin{equation}
	\hat\theta = \arcsin(\angle z/\pi)
\end{equation}
where $z$ denotes a selected signal root. The network parameters are optimized using the root mean squared periodic error (RMSPE) as the loss function $\Loss$.

\section{Bottleneck of the Spectral Decomposition}
\label{sec:gargalo}
\subsection{EVD Mechanics and Computational Complexity}
\label{sec:evd}

The eigendecomposition (EVD) of the surrogate covariance matrix $\mRz$ is defined as
\begin{equation}
	\label{eq:evd}
	\mRz=\mU\mLam\mU\herm,
\end{equation}

where $\mU \in\mathbb{C}^{N\times N}$ is unitary, with orthonormal eigenvectors as its columns, and
$\mLam=\diag(\lambda_1,\ldots,\lambda_N)$ contains the real eigenvalues. Numerical standard eigensolver implementations of EVD of dense Hermitian matrices, first reduce the matrix to tridiagonal form and subsequently solve the resulting tridiagonal eigenvalue problem~\cite{anderson1999lapack,golub2013,demmel1997}.

The reduction of the dense Hermitian matrix $\mRz$ to tridiagonal form can be performed using a sequence of unitary Householder transformations~\cite{householder1958}. The complete transformation can be expressed as
\begin{equation}
	\label{eq:householder}
	\mT = (\mathbf{H}_{N-2}\cdots\mathbf{H}_1)\,\mRz\,(\mathbf{H}_1\cdots\mathbf{H}_{N-2})\herm .
\end{equation}
where $\mathbf{H} \in\mathbb{C}^{N\times N}$ is the Householder reflector and $\mT$ is Hermitian and tridiagonal, containing nonzero entries only on the main diagonal and the two adjacent diagonals. The tridiagonalization stage has cubic computational complexity $\bigO(N^3)$ and constitutes the dominant cost of the EVD for dense matrices. Once the matrix has been reduced to tridiagonal form, the eigenvalue problem becomes substantially less expensive because $\mT$ contains only $3N-2$ nonzero entries. Modern eigensolvers such as the divide-and-conquer algorithm~\cite{cuppen1981} recursively partitions the tridiagonal eigenvalue problem into smaller subproblems and solves them independently, with considerable low computational cost.

\subsubsection{Overall Computational Cost}

the EVD of an $N\times N$ Hermitian matrix therefore has computational complexity $\bigO(N^3)$, dominated by the tridiagonalization stage~\cite{parlett1998,golub2013,demmel1997}. The leading constant is on the order of $9$ operations per unit of $N^3$ for the real symmetric case and approximately four times larger for the complex Hermitian case, reflecting the additional cost of complex floating-point operations. It is worth noting that this cubic scaling arises from the orthogonal similarity transformations used to compute the EVD. The proposed approach eliminates the need for the complete spectral decomposition.

\subsubsection{Impact on SubspaceNet}

assuming a reference scenario of $N=8$, the EVD takes approximately $4$--$6$~ms per call and is therefore negligible. For $N=1024$, the runtime increases to approximately $1.7$~s, while for $N=2048$, it reaches approximately $15.7$~s (measured values reported in Section~\ref{sec:ganho}). The cubic scaling increases the computational cost by a factor of eight each time $N$ is doubled. Since the differentiable Root-MUSIC triggers the EVD (and its backpropagation) \emph{once per batch sample, per epoch}, this cost accumulates throughout the training process. The proposed alternative reduces this cost to $\bigO(mN^2)$, with $m$ kept fixed; a detailed analysis is presented in Section~\ref{sec:custo}.

\subsection{Instability of Backpropagation Through the Complex EVD}
\label{sec:instabilidade}

The CNN-based subspace method (SubspaceNet) is trained \emph{end-to-end}. The loss (RMSPE) is evaluated at the output of Root-MUSIC, \emph{after} the eigendecomposition. To update the weights of the CNN that produces $\mRz$, the loss gradient must propagate backward through the EVD, and its Jacobian must be applied at every training step. This gradient computation is particularly fragile~\cite{parlett1998,golub2013,7410696,stewart1990}. According the \emph{perturbation theory} of eigenvectors~\cite{stewart1990}, when the surrogate covariance matrix $\mRz$ is perturbed by an infinitesimal increment $d\mRz$, the first-order variation of the $i$-th eigenvector can be expressed as a linear combination of the \emph{other} eigenvectors:

\begin{equation}
	\label{eq:perturbacao}
	d\vu_i = \sum_{j\neq i} \frac{\vu_j\herm\,(d\mRz)\,\vu_i}{\lambda_i - \lambda_j}\ \vu_j
\end{equation}

The weight associated with each direction contains the spectral gap $\lambda_i-\lambda_j$ in the denominator and when two eigenvalues are close to each other, even a small perturbation in $\mRz$ can cause a large change in the corresponding eigenvectors. Thus, the eigenvectors become \emph{ill-conditioned} as the spectral gaps become small. During backpropagation, this sensitivity and the factor $1/(\lambda_i-\lambda_j)$ appears in the Jacobian of the EVD. Thus, two closely spaced eigenvalues may lead to excessively large gradients and numerical overflow.

The diagonally loaded construction of the surrogate $\mathbf R_z$ (\ref{eq:surrogate}) clusters the $N-M$ noise eigenvalues around $\lambda\approx\epsilon$. At the beginning of training, $\mK$ has a low effective rank, with spectral gaps approaching zero within the noise subspace. Consequently, any gradient involving pairs of noise eigenvectors becomes highly sensitive to perturbations in $\mRz$. Also, under \emph{coherent sources or low SNR}, even the signal eigenvalues may become close to one another or approach the noise floor. Both situations can naturally occur during training.

Another source of ambiguity arises from the phase indeterminacy of eigenvectors of a complex Hermitian matrix. Each eigenvector is defined only up to an arbitrary phase factor. Thus, if $\vu_i$ is an eigenvector, then $e^{\ju\varphi_i}\vu_i$ is an equally valid eigenvector for any phase $\varphi_i$. Consequently, if
$\mRz=\mU\mLam\mU\herm$ is an eigendecomposition of $\mRz$, then
\begin{equation}
	\label{eq:fase}
	\mRz
	=
	\mU\mLam\mU\herm
	=
	(\mU\mathbf{D})\mLam(\mU\mathbf{D})\herm,
\end{equation}
where
$\mathbf{D}=\diag(e^{\ju\varphi_1},\dots,e^{\ju\varphi_N})$ is a diagonal unitary matrix. Therefore, the eigenvector directions are uniquely determined, but their individual phases are not.

Although these phase factors do not affect the matrix, they introduce an ambiguity in the representation of the eigenvectors. Consequently, small numerical perturbations can lead to different evaluations of the EVD, causing the computed eigenvectors to change abruptly, which can make the gradient of the loss function with respect to $\mU$ inconsistent across training batches.

The Lanczos-based approach described in Section~\ref{sec:proposta} addresses both sources of instability. The $M$ dominant eigenvalues, from the signal subspace, are separated from the clustered noise eigenvalues. The corresponding spectral gaps are therefore substantially larger, reducing the sensitivity of the eigenvectors to perturbations and improving gradient stability. The noise projector is then obtained from the orthogonal complement of the estimated signal subspace, without explicitly computing the eigenvectors associated with the nearly degenerate noise eigenvalues.

Finally, the eigenvalue problem is solved on a small real symmetric tridiagonal matrix. Its eigenvectors are real and therefore exhibit only sign ambiguity, rather than the continuous phase ambiguity of complex eigenvectors.

\section{Proposed Method}
\label{sec:proposta}

\label{sec:ideia}

Root-MUSIC does not requires the noise eigenvalues, but the \emph{orthogonal projector} onto the noise subspace,
\begin{equation}
	\label{eq:projetor}
	\mPn = \mUn\mUn\herm ,
\end{equation}
which can be written as the orthogonal complement of the signal subspace:
\begin{equation}
	\label{eq:complemento}
	\mPn = \mUn\mUn\herm = \mI - \mUs\mUs\herm .
\end{equation}

Thus, the proposed approach replaces, exclusively within the differentiable Root-MUSIC stage, the full eigendecomposition with a Lanczos tridiagonalization~\cite{lanczos1950}, a Krylov subspace method that constructs the dominant subspace using only matrix--vector products. Neither $\mUn$ nor the $N-M$ noise eigenvalues need to be computed; it is sufficient to determine the signal subspace $\mUs$, whose dimension $M$ is much smaller than $N$.

The method is designed to  reduce the computational cost from $\bigO(N^3)$ for the full EVD to $\bigO(mN^2)$ for Lanczos with fixed $m$, where $m\ll N$, making subspace extraction feasible for large arrays with tens to thousands of sensors, such as those encountered in massive MIMO systems, for which the full EVD becomes prohibitively expensive.

Also, the proposed approach replace backpropagation through the complex-valued EVD, which is unstable, as  discussed in Section~\ref{sec:instabilidade}, with backpropagation through the eigendecomposition of a small real symmetric matrix, whose only ambiguity is a sign $(\pm1)$ and is therefore much better behaved than the continuous phase ambiguity $e^{\ju\varphi}$. Finally, the accuracy is preserved, ensuring that the replacement is mathematically transparent, such that the modified model computes the same estimator as the original one, rather than an approximation.

\subsection{Lanczos Tridiagonalization}
\label{sec:lanczos}

The Lanczos method~\cite{lanczos1950,saad2003,bai2000templates,golub2013} iteratively constructs an orthonormal basis $\mV=[\vv_1,\dots,\vv_m]$. Here, $\vv_1$ is the initial unit-norm vector used to generate the Krylov subspace:
\begin{equation}
	\label{eq:krylov}
	\Kry_m(\mRz,\vv_1) = \Span\{\vv_1,\ \mRz\vv_1,\ \dots,\ \mRz^{m-1}\vv_1\}
\end{equation}
in which the surrogate covariance matrix is represented by a tridiagonal matrix,
$\mRz\mV\approx\mV\mT$, where $\mT$ is a real symmetric $m\times m$ matrix with $m\ll N$.

At the end of the procedure, the tridiagonal matrix is diagonalized as $\mTm = \mQ\,\mLam_m\,\mQ\trans$, where the columns of $\mQ$ are the eigenvectors of $\mTm$ and $\mLam_m$ holds its eigenvalues. This costs only $\bigO(m^3)$, which is inexpensive because $m$ is small and fixed.

\subsection{Computational Cost: From $\bigO(N^3)$ to $\bigO(mN^2)$}
\label{sec:custo}

The Lanczos algorithm constructs the dominant subspace by performing only $m$ iterations, with each iteration dominated by a single operation: the matrix--vector product $\vr=\mRz\vv$.
$\mRz$ is a dense $N\times N$ matrix, while $\vv$ is an $N\times1$ vector. Each of the $N$ components $r_i$ of the resulting vector is given by the inner product between one row of $\mRz$ and $\vv$:
\begin{equation}
	\label{eq}
	r_i=\sum_{j=1}^{N}[\mRz]_{ij}v_j,
	\qquad i=1,\dots,N.
\end{equation}
Each component requires $N$ multiplications and $N-1$ additions, resulting in $\bigO(N)$ operations. Since $N$ components must be computed, the complete matrix--vector product requires $N\times\bigO(N)=\bigO(N^2)$ operations.
This contrasts with the EVD, for which the Householder reduction applies $N-2$ transformations, each costing $\bigO(N^2)$, resulting in $\bigO(N^3)$ operations for the tridiagonalization alone.

The Lanczos loop repeats the matrix--vector product $m$ times, once per iteration. Multiplying the per-iteration cost by the number of iterations and adding the final diagonalization of the $m\times m$ tridiagonal matrix gives:
\begin{equation}
	\label{eq:custo_total}
	\underbrace{m\times\bigO(N^2)}_{\text{produtos matriz--vetor}}
	\;+\; \underbrace{\bigO(m^3)}_{\text{diagonalization of }\mT} \;=\; \bigO(mN^2),
\end{equation}
which holds whenever $m\ll N$. The $\bigO(m^3)$ term is negligible compared with $\bigO(mN^2)$: in the \emph{benchmark}, with $m=12$ and $N$ up to $2048$, we have $m^3\approx1.7\times10^3$ compared with
$mN^2\approx5\times10^7$, a difference of approximately four orders of magnitude~\cite{saad2003,bai2000templates,golub2013}.

Explicitly forming the projector $\mF=\mI-\mUs\mUs\herm$ would require storing and operating on a \emph{dense} $N\times N$ matrix, resulting in $\bigO(N^2)$ memory and computational cost per angle evaluation. This is avoided because the proposed method stores only the signal subspace basis $\mUs$, an $N\times M$ matrix with $M\ll N$ (in the reference scenario, $8\times3$), and never explicitly forms $\mF$.

For each angle, considering $\va\equiv\va(\theta)$ the steering vector to angle $\theta$, the MUSIC pseudospectrum can be evaluated using the identity:
\begin{equation}
	\label{eq:naomaterializa}
	\va\herm\big(\mI-\mUs\mUs\herm\big)\va=
	||\va||_2^2-||\mUs\herm\va||_2^2,
\end{equation}
where vector $\mathbf U_s^H\mathbf a(\theta) \in \mathbb{C}^{M}$ represents the coordinates of the steering vector $\mathbf a(\theta)$ projected onto the $M$-dimensional signal subspace~\cite{vantrees2002}. The computational cost per angle is therefore reduced from $\bigO(N^2)$ to $\bigO(NM)$.

Thus, in addition to avoiding the full EVD, the proposed approach also avoids explicitly forming and inverting any $N\times N$ matrix.

\subsection{Structure of the Proposed Method}
\label{sec:inferencia}

Fig.~\ref{fig:inferencia} illustrates the original architecture, which uses a CNN to infer a surrogate covariance matrix, alongside the architecture incorporating the proposed method.

\begin{figure}[!t]
	\centering
	\begin{tikzpicture}[
			font=\scriptsize, >=Stealth,
			blk/.style={draw, rounded corners=2pt, align=center, inner sep=3pt,
					minimum height=8mm, text width=36mm},
			evd/.style={blk, fill=evdc!10, draw=evdc!85!black},
			lzs/.style={blk, fill=lzc!12, draw=lzc!85!black},
			com/.style={blk, text width=80mm, fill=gray!8, draw=gray!70!black},
			arr/.style={->, semithick}
		]
		\node[com] (in) at (0,0) {Snapshots $\mathbf{X}\in\mathbb{C}^{N\times T}$ $\;\rightarrow\;$ CNN (SubspaceNet [6])\\[1pt]
			$\rightarrow\;$ surrogate covariance $\mathbf{R}_z\in\mathbb{C}^{N\times N}$, Eq.~(2)};

		\node[font=\scriptsize\bfseries, text=evdc!85!black] (h1) at (-2.1,-1.0) {Original [6]};
		\node[font=\scriptsize\bfseries, text=lzc!85!black]  (h2) at ( 2.1,-1.0) {Proposed};

		\node[evd, below=1.5mm of h1, minimum height=13.5mm] (e1) {\textbf{Full complex EVD}\\ all $N$ eigenpairs, $\mathcal{O}(N^3)$\\[1pt]
		{\tiny\itshape\color{black!65} backprop: $1/(\lambda_i-\lambda_j)$ for all pairs; phase ambiguity}};
		\node[evd, below=3mm of e1, minimum height=11mm] (e2) {Noise subspace\\ $\mathbf{U}_n\in\mathbb{C}^{N\times(N-M)}$};
		\node[evd, below=3mm of e2] (e3) {$\mathbf{F}=\mathbf{U}_n\mathbf{U}_n^{\mathsf H}$};

		\node[lzs, below=1.5mm of h2, minimum height=13.5mm] (l1) {\textbf{Lanczos}\\ $m\ll N$ products $\mathbf{R}_z\mathbf{v}$, $\mathcal{O}(mN^2)$};
		\node[lzs, below=3mm of l1, minimum height=11mm] (l2) {eig of real tridiagonal\\ $\mathbf{T}_m\in\mathbb{R}^{m\times m}$, $\mathcal{O}(m^3)$\\[1pt]
		{\tiny\itshape\color{black!65} backprop: real $m\times m$ eig; sign ambiguity only}};
		\node[lzs, below=3mm of l2] (l3) {$\mathbf{U}_s\in\mathbb{C}^{N\times M}$,\; $\mathbf{F}=\mathbf{I}-\mathbf{U}_s\mathbf{U}_s^{\mathsf H}$};

		\path (e3.south) -- (l3.south) coordinate[midway] (mid);
		\node[com, text width=50mm, anchor=north] (rm) at ($(mid)+(0,-0.55)$) {Root-MUSIC (same in both paths) $\;\rightarrow\;\hat{\boldsymbol\theta}$};

		\draw[arr] ($(in.south)+(-2.1,0)$) -- (h1.north);
		\draw[arr] ($(in.south)+( 2.1,0)$) -- (h2.north);
		\draw[arr, evdc!85!black] (e1) -- (e2);  \draw[arr, evdc!85!black] (e2) -- (e3);
		\draw[arr, lzc!85!black]  (l1) -- (l2);  \draw[arr, lzc!85!black]  (l2) -- (l3);
		\draw[arr, evdc!85!black] (e3.south) |- ($(rm.north)+(-1.2,0.25)$) -- ($(rm.north)+(-1.2,0)$);
		\draw[arr, lzc!85!black]  (l3.south) |- ($(rm.north)+( 1.2,0.25)$) -- ($(rm.north)+( 1.2,0)$);
	\end{tikzpicture}
	\caption{Spectral-extraction stage of SubspaceNet [6] (left) and of the proposed method (right). Both paths receive the same learned surrogate covariance $\mathbf{R}_z$ and feed the same Root-MUSIC estimator. The original path computes all $N$ eigenpairs of $\mathbf{R}_z$ with a full complex EVD, at $\mathcal{O}(N^3)$ cost, and builds the projector from the $N-M$ noise eigenvectors. The proposed path runs $m\ll N$ Lanczos iterations, each dominated by one product $\mathbf{R}_z\mathbf{v}$, reduces $\mathbf{R}_z$ to a real symmetric $m\times m$ tridiagonal matrix $\mathbf{T}_m$, and obtains the projector from the $M$-dimensional signal subspace as $\mathbf{F}=\mathbf{I}-\mathbf{U}_s\mathbf{U}_s^{\mathsf H}$. The italic lines give the decomposition through which the loss gradient is backpropagated during training and its ambiguity.}
	\label{fig:inferencia}
\end{figure}
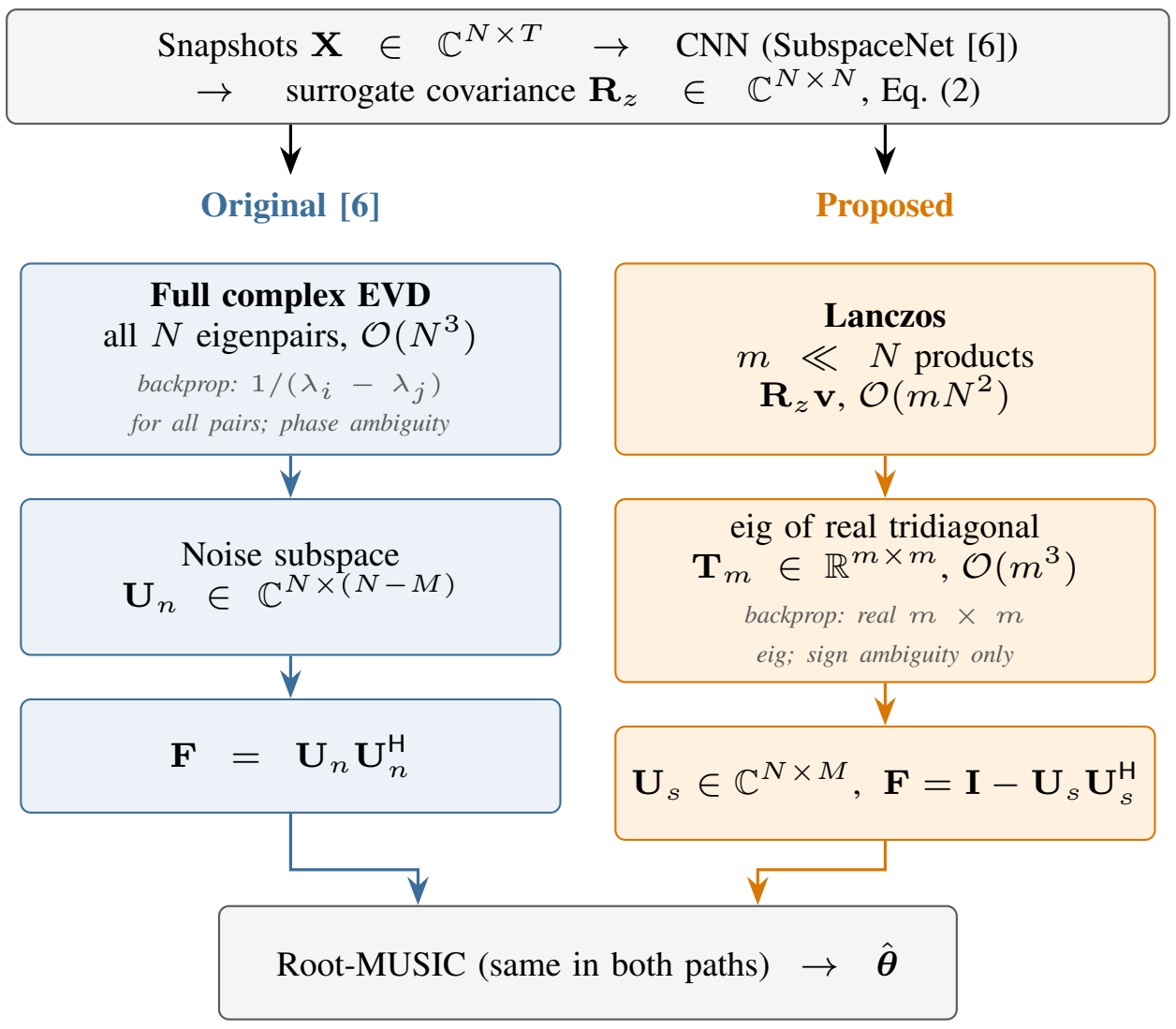
The original routine includes the full EVD step, corresponding to \emph{a single} call to a highly optimized linear algebra procedure.
The second estage of proposed path is the implementation of an iterative Lanczos procedure. Thus, instead of a single optimized call, the proposed path requires multiple tensor operations, introducing additional computational overhead.

\section{Results}
\label{sec:resultados}

\subsection{Experimental Setup}
\label{sec:setup}

Unless otherwise stated, we adopt the reference scenario from~\cite{10750453} (Fig.~\ref{fig:setup}):
$N=8$ sensors, $M=3$ sources, $T=100$ \emph{snapshots}, an SNR of $10$~dB,
and noncoherent sources.
This section compares the original SubspaceNet~\cite{10750453}, referred to as the EVD model, with the proposed model, referred to as the Lanczos model, which is identical except that the full EVD in the differentiable Root-MUSIC is replaced by the Lanczos procedure of Section~IV.

The two models were trained under strictly identical
conditions, including the same random seed, identical initial weights, and the
same batch ordering, using the Adam optimizer~\cite{kingma2015adam} and
PyTorch~\cite{paszke2019pytorch}. The scalability benchmarks use a fixed
Krylov subspace dimension of $m=12$, with $N$ ranging from $64$ to $2048$.

\subsection{Statistical Equivalence with the Original Model}

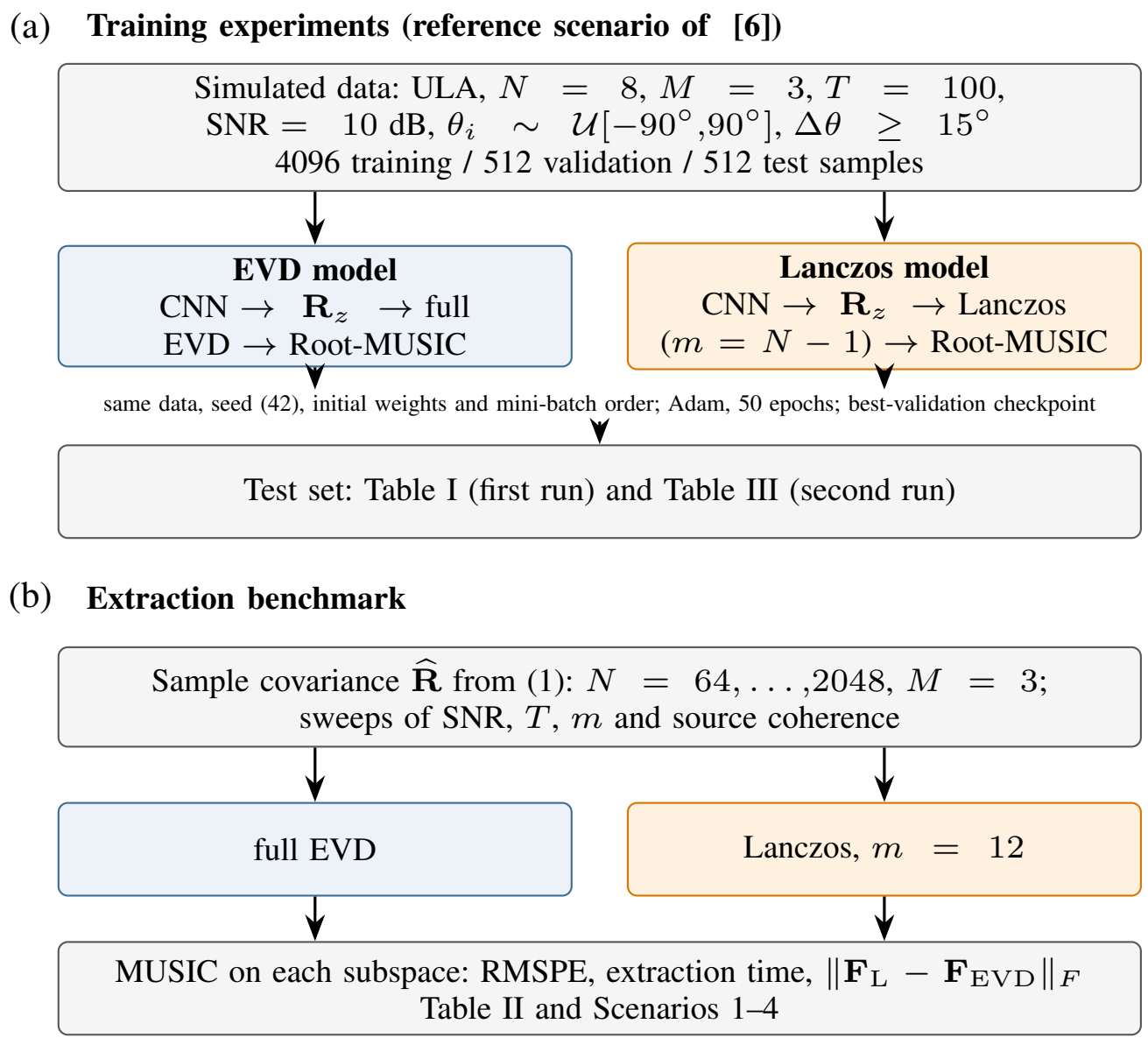
\begin{figure}[!t]
	\centering
	\begin{tikzpicture}[
			font=\scriptsize, >=Stealth,
			blk/.style={draw, rounded corners=2pt, align=center, inner sep=2.5pt, minimum height=7mm},
			com/.style={blk, text width=80mm, fill=gray!8, draw=gray!70!black},
			evd/.style={blk, text width=37mm, fill=evdc!10, draw=evdc!85!black},
			lzs/.style={blk, text width=37mm, fill=lzc!12, draw=lzc!85!black},
			arr/.style={->, semithick}]
		\node[font=\footnotesize] at (-4.3,0.75) {(a)};
		\node[font=\scriptsize\bfseries, anchor=west] at (-4.0,0.75) {Training experiments (reference scenario of ~\cite{10750453})};
		\node[com] (data) at (0,0) {Simulated data: ULA, $N=8$, $M=3$, $T=100$, SNR $=10$ dB,
			$\theta_i\sim\mathcal{U}[-90^\circ,90^\circ]$, $\Delta\theta\ge15^\circ$\\
			4096 training / 512 validation / 512 test samples};
		\node[evd] (me) at (-2.15,-1.35) {\textbf{EVD model}\\ CNN $\rightarrow\mathbf{R}_z\rightarrow$ full EVD $\rightarrow$ Root-MUSIC};
		\node[lzs] (ml) at ( 2.15,-1.35) {\textbf{Lanczos model}\\ CNN $\rightarrow\mathbf{R}_z\rightarrow$ Lanczos ($m=N-1$) $\rightarrow$ Root-MUSIC};
		\node[font=\tiny, align=center, text width=80mm, fill=white, inner sep=1pt] (same) at (0,-2.1)
		{same data, seed (42), initial weights and mini-batch order; Adam, 50 epochs; best-validation checkpoint};
		\node[com] (test) at (0,-2.75) {Test set: Table I (first run) and Table III (second run)};
		\draw[arr] (data.south) ++(-2.15,0) -- (me.north);
		\draw[arr] (data.south) ++( 2.15,0) -- (ml.north);
		\draw[arr] (me.south) -- (me.south |- same.north);
		\draw[arr] (ml.south) -- (ml.south |- same.north);
		\draw[arr] (same.south) -- (test.north);
		\begin{scope}[yshift=-4.3cm]
			\node[font=\footnotesize] at (-4.3,0.75) {(b)};
			\node[font=\scriptsize\bfseries, anchor=west] at (-4.0,0.75) {Extraction benchmark };
			\node[com] (cov) at (0,0) {Sample covariance $\widehat{\mathbf{R}}$ from (1): $N=64,\ldots,2048$, $M=3$;\\
				sweeps of SNR, $T$, $m$ and source coherence};
			\node[evd] (be) at (-2.15,-1.15) {full EVD};
			\node[lzs] (bl) at ( 2.15,-1.15) {Lanczos, $m=12$};
			\node[com] (out) at (0,-2.2) {MUSIC on each subspace: RMSPE, extraction time, $\|\mathbf{F}_{\mathrm{L}}-\mathbf{F}_{\mathrm{EVD}}\|_F$\\
			Table II and Scenarios 1--4};
			\draw[arr] (cov.south) ++(-2.15,0) -- (be.north);
			\draw[arr] (cov.south) ++( 2.15,0) -- (bl.north);
			\draw[arr] (be.south) -- (be.south |- out.north);
			\draw[arr] (bl.south) -- (bl.south |- out.north);
		\end{scope}
	\end{tikzpicture}

	\caption{Experimental setup. (a) Training experiments in the reference scenario: the EVD and Lanczos models are trained with the same data, seed, initial weights and mini-batch order, and evaluated on the same test set. (b) Extraction benchmark: without the CNN, the full EVD and the Lanczos procedure ($m=12$) are applied to the same sample covariance matrices for $N=64,\ldots,2048$ and compared in accuracy, extraction time and projector distance.}
	\label{fig:setup}
\end{figure}

\label{sec:precisao}

Table~\ref{tab:precisao} compares the two models on the test set of the reference scenario, formed by 512 samples not used in training. The values are the mean and the median, over the test samples, of the RMSPE between the true and the estimated DoAs, obtained with the weights of the epoch with the lowest validation RMSPE, which is also reported (best validation). The two models show essentially identical performance.

\begin{table}[!h]
	\caption{Comparison on the test set of the reference scenario}
	\label{tab:precisao}
	\centering\footnotesize
	\setlength{\tabcolsep}{4pt}
	\begin{tabular}{@{}lccc@{}}
		\toprule
		Metric          & Original (EVD) & Lanczos      & Difference             \\
		\midrule
		Mean test RMSPE & $2.04^\circ$   & $2.04^\circ$ & $<0.01^\circ$ (0.05\%) \\
		Median RMSPE    & $0.54^\circ$   & $0.53^\circ$ & $-0.01^\circ$          \\
		Best validation & $1.90^\circ$   & $1.94^\circ$ & $+0.04^\circ$          \\
		\bottomrule
	\end{tabular}\\[1mm]
	\parbox{\columnwidth}{\scriptsize $N=8$, $M=3$, $T=100$, SNR $=10$~dB, 512 test samples. Mean and median of the per-sample test RMSPE obtained with the weights of the epoch with the lowest validation RMSPE (best validation). Difference $=$ Lanczos $-$ EVD.}
\end{table}


A broader experimental evaluation, presented in Sections V-C and V-E, applies the two subspace-extraction methods (with same signal subspace for $M=3$) to sample covariance matrices of arrays with $N=64$ to $2048$ sensors, in 102 configurations that vary $N$, the SNR, the number of snapshots $T$, the Krylov subspace dimension $m$ and the presence of coherent sources. In all of them, the RMSPE obtained with Lanczos matches that of the EVD to the fourth decimal place, except for direct MUSIC with fully coherent sources (Scenario~4 - section \ref{sec:robustness}, in which neither method estimates the DoAs. Fig.~\ref{fig:doas} illustrates this agreement for one test sample: the estimates of the two models nearly coincide, including the larger error, common to both, for the source at $84.4^\circ$, close to endfire.


\begin{figure}[!htbp]
	\centering
	\includegraphics[width=\columnwidth]{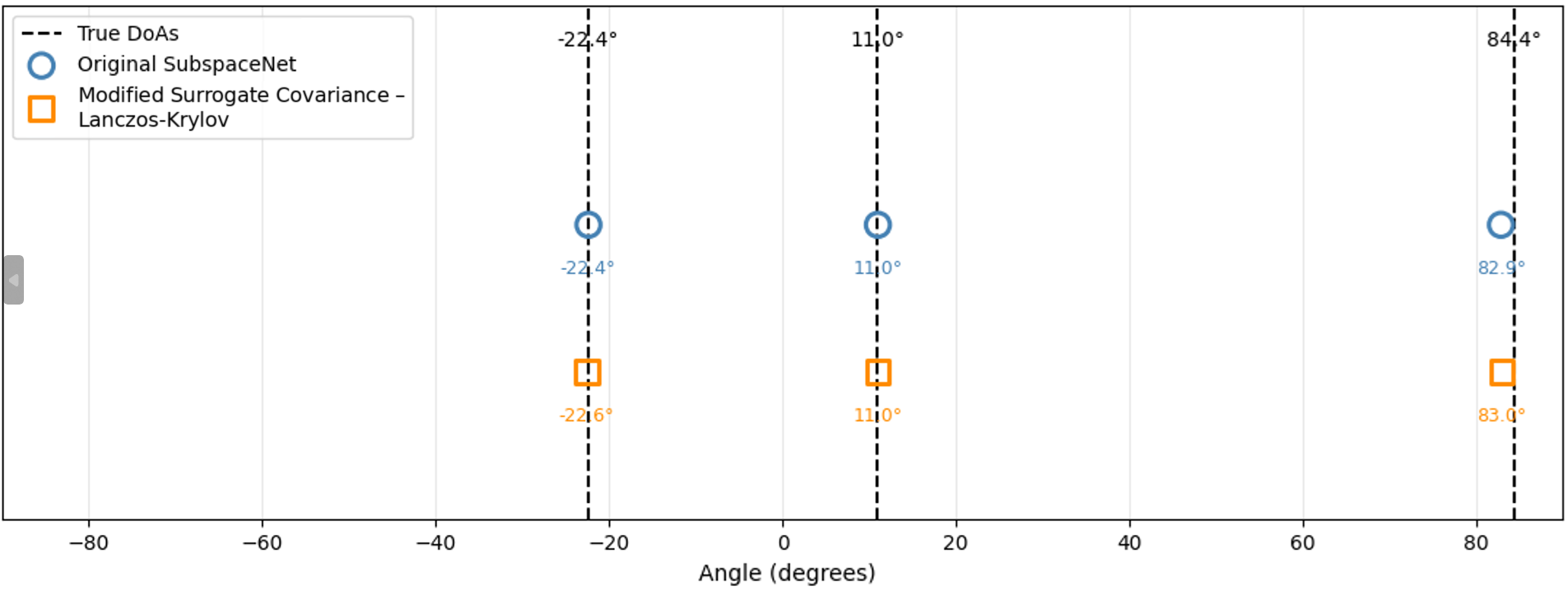}
	\caption{True and estimated DoAs for one test sample of the reference scenario ($N=8$, $M=3$, $T=100$, SNR $=10$~dB). Dashed lines: true DoAs; circles: EVD model; squares: Lanczos model. The vertical position of the markers only separates the two models. }
	\label{fig:doas}
\end{figure}

\subsection{Computational Scalability Gains}
\label{sec:ganho}

In the small reference scenario ($N=8$), the Lanczos approach does not provide a speed advantage. On the contrary, it requires $5.29$~ms compared with $3.81$~ms for the original implementation, making it approximately $39\%$ slower. This behavior is expected and does not contradict the proposed approach: when $m\approx N$, the asymptotic cost $\bigO(mN^2)$ becomes comparable to $\bigO(N^3)$, while the proposed method also performs the spectral extraction through five stages rather than three, thereby incurring the dispatch overhead discussed in Section~\ref{sec:inferencia}.

The computational advantage emerges when $N\gg M$ and $m$ remains fixed, which is precisely the regime encountered in large-scale arrays. A scalability benchmark, with $m=12$ fixed while $N$ increases from $64$ to $2048$, produced the results shown in Table~\ref{tab:escalabilidade}.
\begin{table}[!t]
	\centering
	\caption{Subspace extraction: full EVD versus Lanczos ($m=12$ fixed).}
	\label{tab:escalabilidade}
	\footnotesize
	\setlength{\tabcolsep}{2pt}
	\begin{tabular}{@{}rrrrc@{}}
		\toprule
		$N$              & \textbf{EVD (ms)}    & \textbf{Lanczos (ms)} &
		\textbf{Speedup} & \textbf{Distance}                                                                                 \\
		\midrule
		$64$             & $5.93$--$6.08$       & $4.26$--$4.65$        & ${\sim}1.3\times$        & ${\sim}2\times10^{-15}$ \\
		$256$            & $30.6$--$36.6$       & $1.7$--$1.9$          & ${\sim}16$--$21\times$   & ${\sim}3\times10^{-15}$ \\
		$512$            & $263.8$--$286.9$     & $3.98$--$4.09$        & ${\sim}66$--$70\times$   & ${\sim}4\times10^{-15}$ \\
		$1024$           & $1\,664$--$1\,831$   & $11.5$--$12.7$        & ${\sim}145\times$        & ${\sim}4\times10^{-15}$ \\
		$2048$           & $15\,654$--$17\,497$ & $57.2$--$60.6$        & ${\sim}274$--$289\times$ & ${\sim}4\times10^{-15}$ \\
		\bottomrule
	\end{tabular}\\[1mm]
	\parbox{\columnwidth}{\scriptsize Subspace extraction with the full EVD and with Lanczos ($m=12$ fixed, $M=3$): extraction time, speedup (EVD time / Lanczos time) and distance between the projectors obtained with the two methods. Ranges correspond to different executions.}
\end{table}

The speedup increases consistently with the expected complexity ratio $N^3/N^2=N$. From $N=512$ onward, it approximately doubles whenever $N$ is doubled, confirming the theoretical scaling exponents (approximately $3$ for the EVD and $2$ for Lanczos). At $N=2048$, the full EVD requires approximately $15.7$~s per covariance matrix, whereas the Lanczos method requires approximately $57$~ms (Fig.~\ref{fig:ganho}), representing the difference between a computationally impractical solution and one suitable for real-time continuous-update applications. At all benchmark points, the MUSIC estimates obtained from the two subspaces are numerically indistinguishable, with projector distances on the order of $10^{-15}$.
\begin{figure}[!htbp]
	\centering
	\includegraphics[width=\columnwidth]{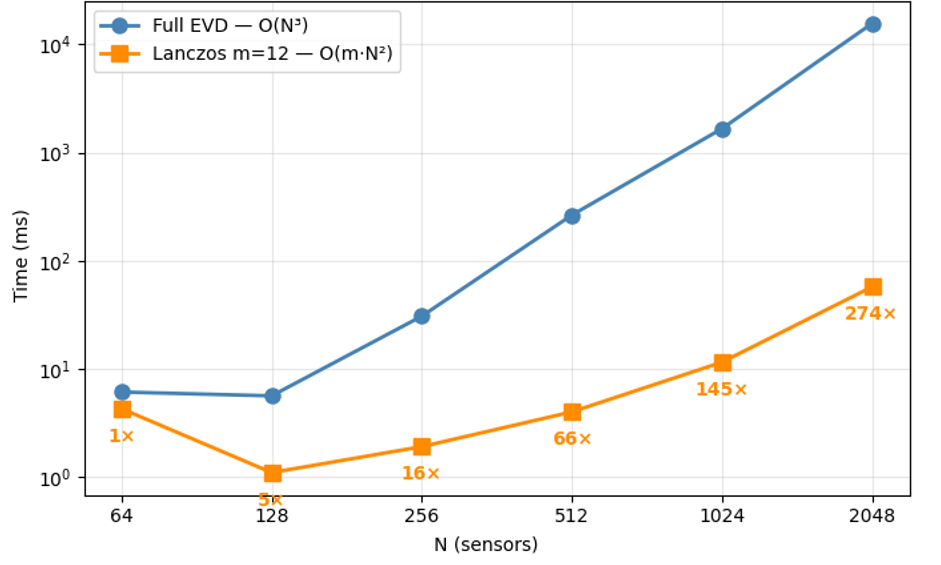}
	\caption{Subspace extraction time as a function of the number of sensors $N$ (logarithmic scales) for the full EVD and for Lanczos with $m=12$ ($M=3$). Labels give the speedup EVD/Lanczos.}
	\label{fig:ganho}
\end{figure}
\subsection{Training Stability}
\label{sec:ensaio}

The complete experimental study included a second run, executed in a new session with the same configuration and the same seed (42) for both models; because the data sets are generated without a fixed seed, this run used newly drawn training, validation and test sets. This run is relevant not because it demonstrates
equivalence, which has already been established, but because it reveals a
qualitative difference in the training behavior of the two gradient
paths, as summarized in Table~\ref{tab:segunda}.

\begin{table}[!htbp]
	\centering
	\caption{Second independent run ($N=8$, same configuration and seed, new data sets).}
	\label{tab:segunda}
	\footnotesize
	\setlength{\tabcolsep}{3.5pt}
	\begin{tabular}{@{}lccc@{}}
		\toprule
		\textbf{Metric} & \textbf{Original (EVD)} & \textbf{Lanczos} & \textbf{Difference}     \\
		\midrule
		Mean test RMSPE & $2.22^\circ$            & $1.84^\circ$     & $-0.38^\circ$ ($-17\%$) \\
		Median RMSPE    & $0.76^\circ$            & $0.63^\circ$     & $-0.13^\circ$ ($-17\%$) \\
		Best validation & $2.04^\circ$            & $1.77^\circ$     & $-0.27^\circ$           \\
		Inference time  & 4.11 ms                 & 5.77 ms          & $+40\%$                 \\
		\bottomrule
	\end{tabular}\\[1mm]
	\parbox{\columnwidth}{\scriptsize Same metrics as Table~\ref{tab:precisao}, obtained in a second session with newly drawn training, validation and test sets. Inference time: mean time per test sample. Difference $=$ Lanczos $-$ EVD.}
\end{table}

At first glance, the Lanczos-based model outperforms the original model in all accuracy metrics, with an improvement of approximately $17\%$. However, this difference should not be interpreted as evidence that the
Lanczos approach is intrinsically more accurate. Rather, it results from a distinct event in the training trajectories. Around epoch~$29$, the original model experienced a catastrophic training collapse: the validation RMSPE increased from approximately $2.5\Odeg$ to approximately $35\Odeg$ and did not fully recover by epoch~$50$.

The final RMSPE of $2.216\Odeg$ does not directly reveal the severity ofvthis collapse because the \emph{checkpoint} mechanism restored the best weights recorded before the event. In contrast, the Lanczos-based model, trained under the same schedule and random seed, passed through its initial transient peaks and subsequently remained stable.
The Lanczos-based model, subjected to the same schedule and the same seed, passes through its initial transient peaks and remains stable throughout training.

A single training run is not sufficient to establish the statistical superiority of one method over the other. The results instead indicate that both approaches achieve comparable accuracy, while the proposed Lanczos-based path exhibits more stable training behavior in this particular run. This observation is consistent with the instability analysis presented in Section~\ref{sec:instabilidade}: an unstable gradient associated with the complex eigendecomposition can perturb the network parameters toward an unfavorable region of the optimization landscape.

\subsection{Robustness}
\label{sec:robustness}

Four additional scenarios stress the proposed approach by varying the SNR,
number of \emph{snapshots}, Krylov subspace dimension, and source coherence.
The main conclusion is that the RMSPE obtained with Lanczos is virtually
identical to that obtained with the EVD, matching to the fourth decimal place
across all configurations. The speedup, in contrast, depends primarily on
$N$ (or on the subarray size) and varies only marginally with the other
factors.

\subsubsection{Scenario 1 --- SNR Sensitivity}

the SNR sweep $\in\{-10,\dots,+20\}$~dB tests an expected hypothesis: since Lanczos relies on eigenvalue separation, and this separation decreases at low SNR, one might expect $m=12$ to fail at $-10$~dB. However, this did not occur in the experiments. The RMSPE obtained with Lanczos is identical to that of the EVD over the entire SNR range, while the speedup ($2\times$, $6\times$, $19\times$, $79\times$, $140\times$, $208\times$) is essentially insensitive to SNR.

The spectral gap \emph{increases with number of sensors $N$}, since the signal eigenvalue associated with a source is on the order of $N\sigma_s^2$, where $\sigma_s^2$ denotes the source power. Thus, the array aperture concentrates the source energy into a dominant eigenvalue, whereas the noise floor remains at approximately the noise power $\sigma_n^2$. Consequently, larger arrays make the Krylov problem \emph{easier}, rather than more difficult, precisely in the regime where the proposed approach is most needed. Figs.~\ref{fig:cen1a} and~\ref{fig:cen1b} show, respectively, the RMSPE \emph{versus} SNR for $N$ ranging from $64$ to $2048$ and the corresponding speedup as a function of $N$.
\begin{figure}[!htbp]
	\centering
	\includegraphics[width=\columnwidth]{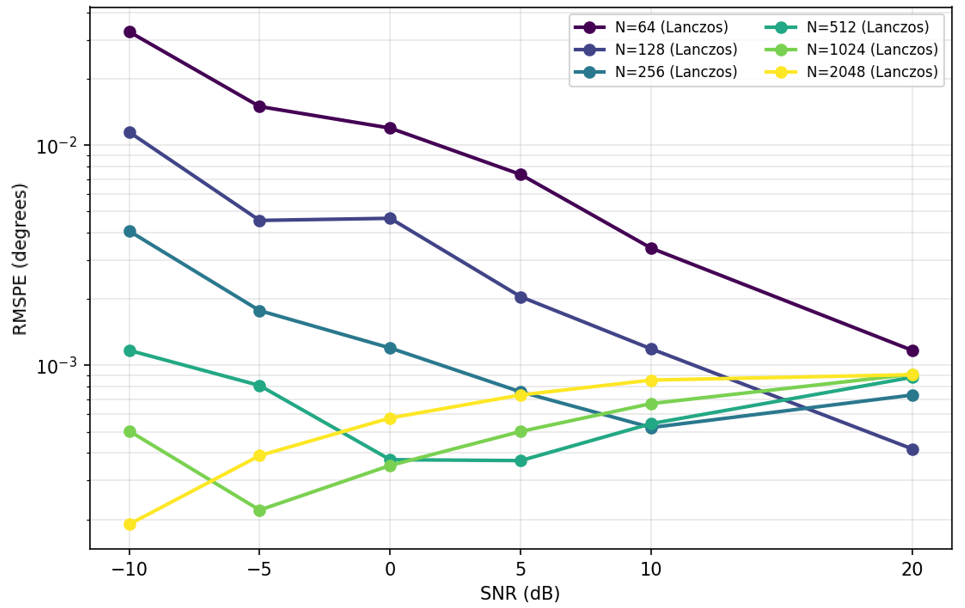}
	\caption{Scenario~1: RMSPE versus SNR for $N=64$ to $2048$, with Lanczos ($m=12$, solid lines) and the full EVD (dashed lines); the curves of the two methods coincide.}
	\label{fig:cen1a}
\end{figure}
\begin{figure}[!htbp]
	\centering
	\includegraphics[width=\columnwidth]{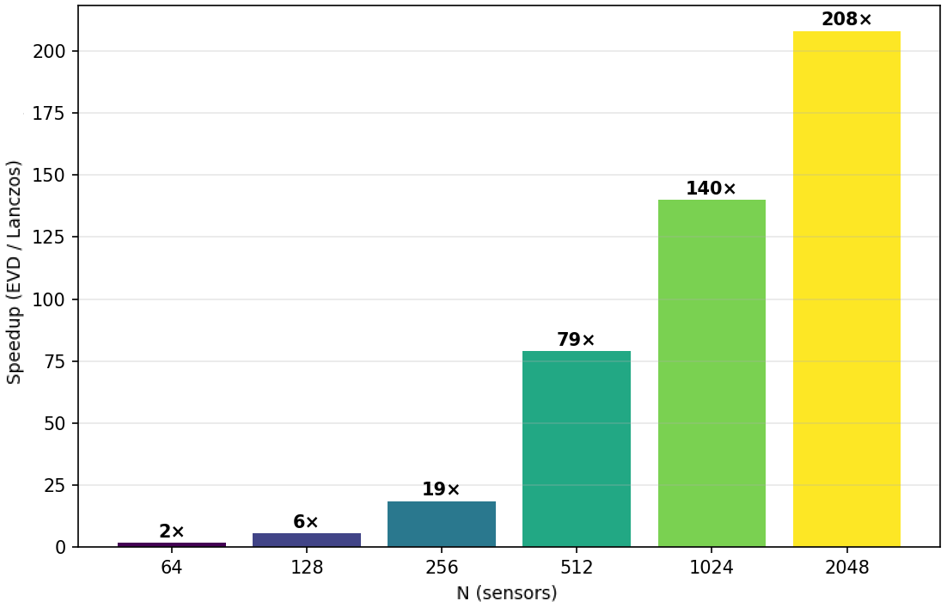}
	\caption{Scenario~1: mean speedup EVD/Lanczos over the SNR values, as a function of $N$.}
	\label{fig:cen1b}
\end{figure}

\subsubsection{Scenario 2 --- Few \emph{Snapshots} ($T\ll N$)}

a sweep over $T\in\{20,50,100,200\}$ was performed, with most cases satisfying $T<N$, for which the sample covariance matrix $\mRhat$ is \emph{singular} (with rank at most $T$). This is a typical operating regime for large-scale arrays (operating $2048$ sensors with $T\geq2048$ \emph{snapshots} is impractical in a dynamic environment). When $T<N$, $\mRhat$ has $N-T$ exactly zero eigenvalues, resulting in a large eigenvalue cluster (for $N=2048$ and $T=20$, there are $2028$ zero eigenvalues).

Direct extraction of the noise subspace would therefore be severely
challenging. However, the signal subspace remains well defined as long
as $T>M$, and the Lanczos procedure can capture it without being affected
by the singularity of the remainder of the spectrum. The result is
accuracy identical to that of the EVD even for $T=20$, with speedups of
approximately $180\times$ for $N=2048$. Figs.~\ref{fig:snapa} and~\ref{fig:snapb} illustrate Scenario 2. Even when $T\ll N$ and $\mRhat$ is singular, the Krylov-based signal subspace remains accurate, while the speedup follows the expected scaling with
$N$, reaching approximately $180\times$ for $N=2048$.

\begin{figure}[!htbp]
	\centering
	\includegraphics[width=\columnwidth]{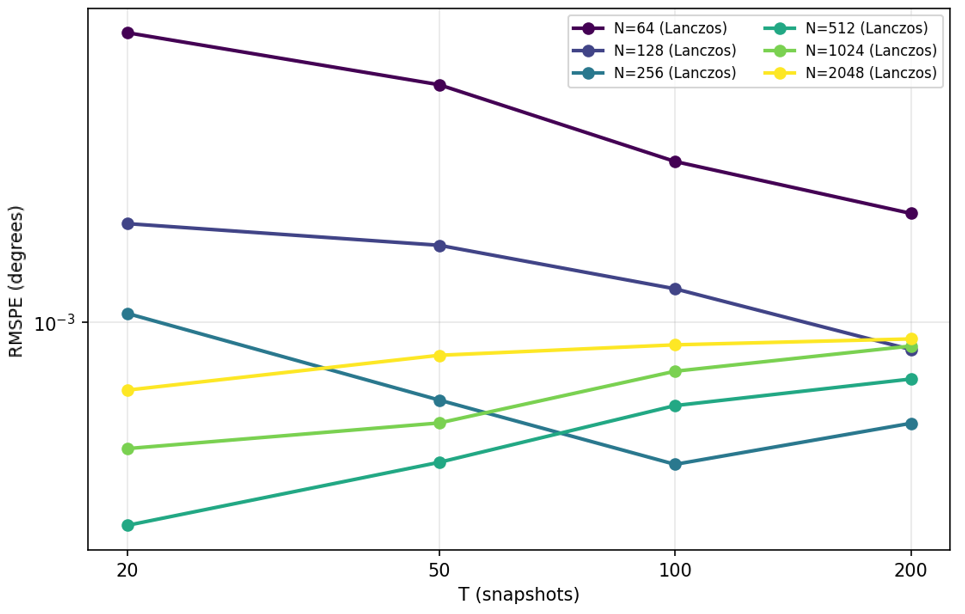}
	\caption{Scenario~2: RMSPE versus the number of snapshots $T$ for $N=64$ to $2048$, with Lanczos ($m=12$, solid lines) and the full EVD (dashed lines). For $T<N$ the sample covariance matrix is singular; the curves of the two methods coincide.}
	\label{fig:snapa}
\end{figure}
\begin{figure}[!htbp]
	\centering
	\includegraphics[width=\columnwidth]{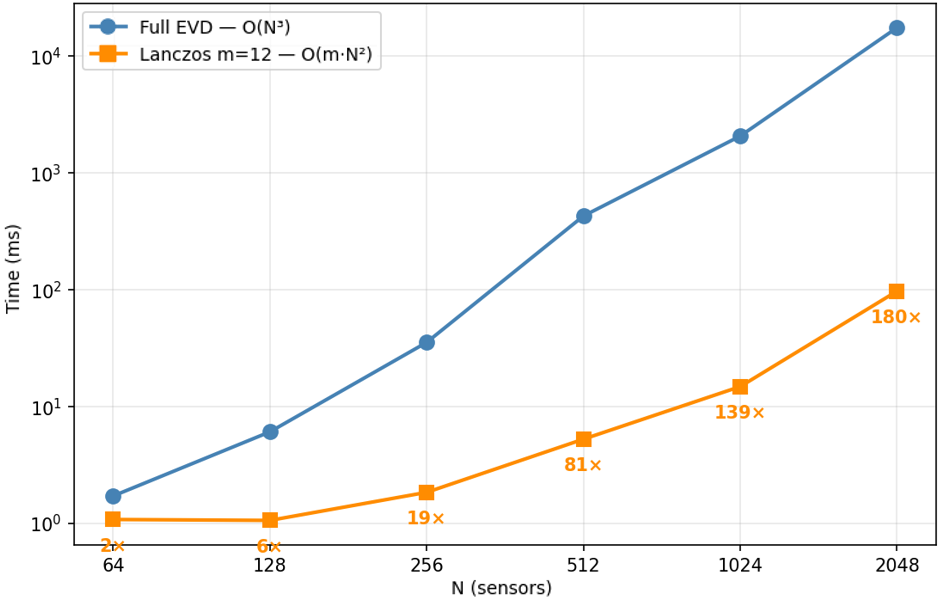}
	\caption{Scenario~2: subspace extraction time as a function of $N$ for the full EVD and for Lanczos with $m=12$ (logarithmic scales). Labels give the speedup EVD/Lanczos, which reaches $180\times$ for $N=2048$.}
	\label{fig:snapb}
\end{figure}
\subsubsection{Scenario 3 --- Krylov Subspace Dimension}

a sweep over $m\in\{8,12,16,20\}$ was performed. Accuracy saturates already at approximately $m\approx2.7M$: the RMSPE curves remain essentially flat and overlap the EVD reference for all $m\geq8$, indicating that any further increase in $m$ only adds computational cost. The speedup decreases approximately linearly with $m$, as predicted by the $\bigO(mN^2)$ complexity: doubling $m$ approximately halves the speedup.

A practical choice is therefore $m\in[2M+2,\,4M]$, indicating that the proposed method does not require fine-tuning of the Krylov dimension. For $N=64$ and $m=20$, the speedup approaches $1\times$: when $m$ becomes on the order of $N/3$, the computational advantage is largely diminished, which defines the useful operating range of the proposed approach.

Accuracy saturates as soon as $m\ge8$, that is, at $m\approx2{,}7M$: across $m\in\{8,12,16,20\}$ the Lanczos RMSPE is constant and matches the full-EVD reference to the fourth decimal place, any further increase in $m$ is pure cost.

\subsubsection{Scenario 4 --- Coherent Sources}

three closely spaced coherent sources ($-6\Odeg$, $0\Odeg$, and $+7\Odeg$) are considered at SNR of $0$~dB, representing the most challenging scenario in
the evaluation. For fully coherent sources, the source covariance matrix has rank one; consequently, the three-dimensional signal subspace assumed by MUSIC does \emph{not} exist, and conventional MUSIC becomes unreliable, exhibiting errors of several tens of degrees for $N\ge1024$.

It is worth noting that, in this failure regime, EVD and Lanczos produce equally invalid estimates, although the numerical estimates may differ ($57\Odeg$ versus $17\Odeg$ for $N=2048$). Such a comparison is \emph{statistically meaningless}, since neither method is correctly estimating the three DoAs. Spatial smoothing, originally proposed by Evans \emph{etal.}~\cite{evans1981} and analyzed by Shan, Wax, and Kailath~\cite{1164649}, is employed here in its \emph{forward--backward} variant~\cite{17496}, using subarrays of $L=N/2$ sensors. Spatial smoothing restores the rank of the signal subspace, and the resulting smoothing $+$ MUSIC \emph{pipeline} achieves RMSPE values between $0{,}0009\Odeg$ and $0{,}0095\Odeg$ for
\emph{all} values of $N$, with EVD and Lanczos agreeing to the fourth decimal place.

Moreover, the computational gain is preserved after spatial smoothing. For the resulting $L\times L$ covariance matrix, the same scaling law applies, now with respect to $L$, reaching a speedup of $112\times$ for $L=1024$.
Thus, robustness and scalability are complementary rather than competing properties. Figs.~\ref{fig:cen4a} and~\ref{fig:cen4b} illustrate the results of scenario 4.

\begin{figure}[!htbp]
	\centering
	\includegraphics[width=\columnwidth]{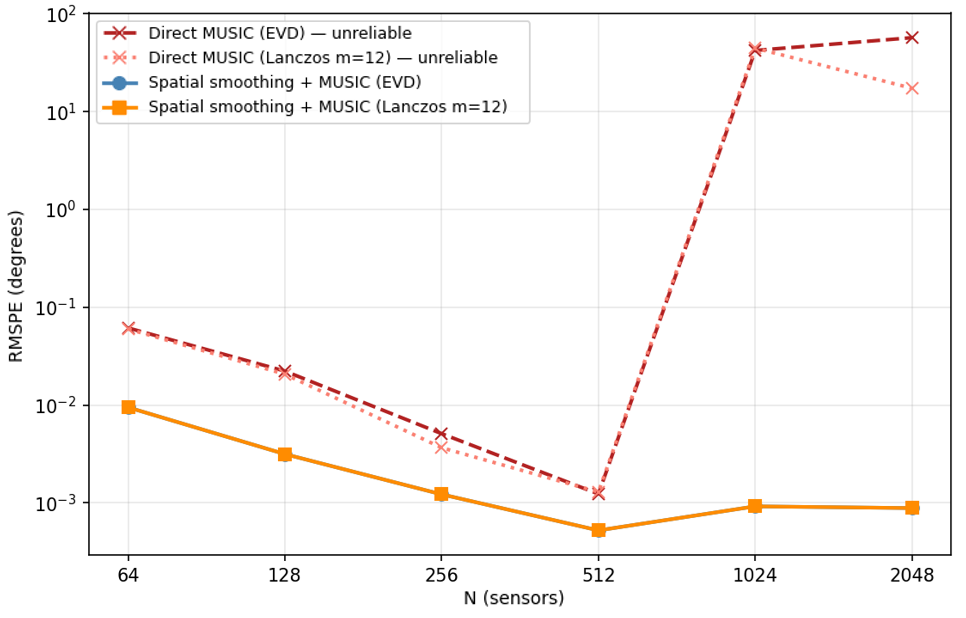}
	\caption{Scenario~4: RMSPE versus $N$ for three coherent sources at $-6^\circ$, $0^\circ$ and $7^\circ$ (SNR $=0$~dB). Direct MUSIC (crosses) is unreliable with both EVD and Lanczos; with forward--backward spatial smoothing over subarrays of $L=N/2$ sensors, the EVD (circles) and Lanczos (squares) curves coincide.}
	\label{fig:cen4a}
\end{figure}

\begin{figure}[!htbp]
	\centering
	\includegraphics[width=\columnwidth]{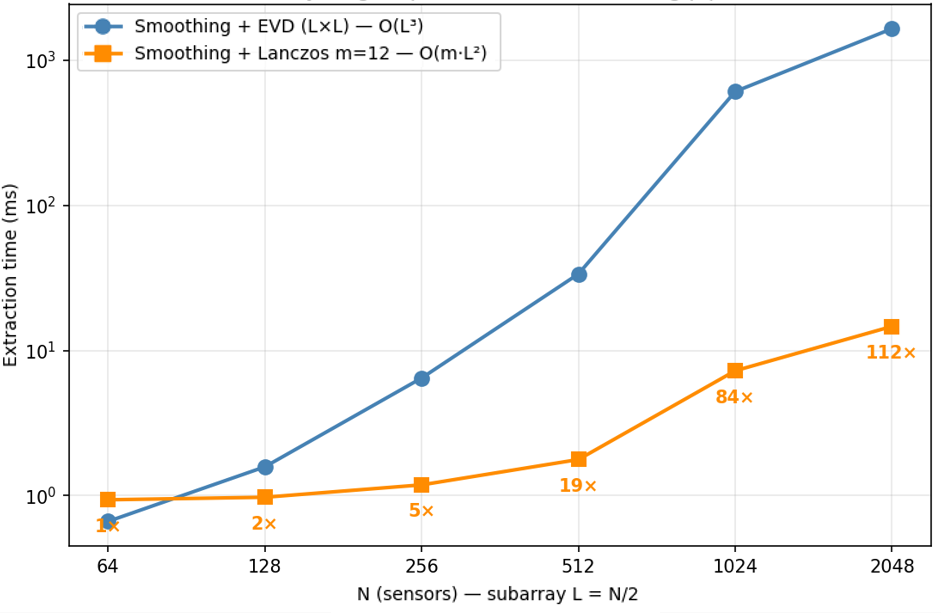}
	\caption{Scenario~4: subspace extraction time for the smoothed $L\times L$ covariance matrix ($L=N/2$) with the full EVD and with Lanczos ($m=12$). Labels give the speedup EVD/Lanczos, which reaches $112\times$ for $L=1024$.}
	\label{fig:cen4b}
\end{figure}
\section{Conclusions}
\label{sec:conclusao}

This work presents the analysis and results of replacing the block that performs a full EVD of a learned surrogate covariance matrix with a CNN-based subspace method (SubspaceNet) before applying the Root-MUSIC algorithm. Although this method replaces the covariance matrix with the learned surrogate, the EVD step still incurs a high computational cost.
In the proposed method, the full EVD is replaced by Krylov subspace and Lanczos tridiagonalization, reducing the computational complexity from $\bigO(N^3)$ to $\bigO(mN^2)$ for fixed $m$. In this scenario, backpropagation runs through the eigendecomposition of a small real symmetric tridiagonal matrix.

Experimental results demonstrate statistical equivalence to the original SubspaceNet throughout the evaluation, which covered more than one hundred configurations varying the number of sensors $N$, the signal-to-noise ratio (SNR), the number of snapshots $T$, the Krylov subspace dimension, and the source coherence. The results show RMSPE values matching those of the original method to four decimal places, speedups of up to $289\times$ for $N=2048$ with no loss of accuracy, and a hybrid diagnostic that isolates and eliminates the effect of inference.

Finally, the article presents a formulation of the Root-MUSIC algorithm in terms of the orthogonal complement of the signal subspace, eliminating the need for the noise subspace and its associated eigenvalues.

\section*{Acknowledgments}

This study was financed in part by the Coordenação de Aperfeiçoamento de Pessoal de Nível Superior - Brasil (CAPES) Finance Code 001, by INCT-NAMITEC (CNPq nº 406193/2022-3) and by FAPERGS (Fundação de Amparo à Pesquisa do Estado do RS), processo 23/2551-0002201-0.

\bibliographystyle{IEEEtran}

\bibliography{DoA}

\end{document}